\documentclass[prc,superscriptaddress]{revtex4}
\usepackage{amssymb,color,graphicx,bm}

\usepackage[caption=false]{subfig}

\definecolor{red}{rgb}{0.8,0,0}
\definecolor{RED}{rgb}{0.8,0,0}
\definecolor{violet}{rgb}{0.4,0,0.4}
\definecolor{green}{rgb}{0,0.5,0.0}
\definecolor{GREEN}{rgb}{0,0.5,0.0}
\definecolor{navy}{rgb}{0.0,0.0,0.6}
\definecolor{orange}{rgb}{0.8,0.2,0.0}
\definecolor{blue}{rgb}{0.3,0.0,0.8}

\newcommand{\gsim}{\raisebox{-.4ex}{$\stackrel{>}{\scriptstyle \sim}$}}

\begin{document}

%\title{Origin of nonlinearity and plausible turbulence by hydromagnetic transient growth 
%dimming magnetorotational instability 
%in accretion disks}
\title{Origin of nonlinearity and plausible turbulence by hydromagnetic transient growth 
in accretion disks: Faster growth rate than magnetorotational instability
}

\author{Sujit Kumar Nath, Banibrata Mukhopadhyay$^*$\\
Department of Physics, Indian Institute of Science, 
Bangalore 560012, India\\ sujitkumar@physics.iisc.ernet.in, bm@physics.iisc.ernet.in\\
%\affiliation{Department of Physics, Indian Institute of Science, 
%       Bangalore 560012, India}
$^*$Corresponding author
}

\begin{abstract}
We investigate the evolution of hydromagnetic perturbations in
% rotating shear flows, in particular 
a small section of accretion disks. It is known that molecular viscosity is negligible 
%to explain the efficiency of infall of matter 
in accretion disks. 
%and hence, astrophysical observation. 
Hence, it has been argued that a mechanism, known as Magnetorotational Instability (MRI), is responsible 
for transporting 
matter in the presence of weak magnetic field. 
%while the magnetic field is ubiquitous in any astrophysical system. 
However, there are some shortcomings, which question effectiveness of MRI.
Now the question arises, whether other hydromagnetic effects, e.g. transient growth (TG), can 
play important role to bring nonlinearity in the system, even at weak magnetic fields. Otherwise, 
whether MRI or TG, which is primarily responsible to reveal nonlinearity to make the flow 
turbulent? Our results prove explicitly that the flows with high Reynolds number ($R_e$), which is the case of 
realistic astrophysical accretion disks, exhibit nonlinearity by TG of perturbation modes 
faster than that by modes producing MRI. 
For a fixed wavevector, MRI dominates over transient effects, only at low $R_e$, lower than 
its value expected to be in astrophysical accretion disks, and low magnetic fields. This seriously questions (overall) 
suasiveness of MRI in astrophysical accretion disks.
\end{abstract}

\maketitle


%\paragraph*{Introduction.$-$} 
\section{Introduction}
Accretion disks are found in 
active galactic nuclei (AGNs), around compact stellar objects in a binary system, around newly formed 
stars etc. (see, e.g., \cite{pringle}). However, the working principle 
of accretion disks still remains enigmatic to us. Due to its inadequacy of 
molecular viscosity, turbulent viscosity has been proposed to explain the transport of matter towards the
central object. This idea is particularly attractive because of 
its high Reynolds number $R_e\gsim 10^{14}$ 
(see, e.g., \cite{bmplb}). However, the Keplerian disks, which are relevant to many astrophysical applications, 
are remarkably Rayleigh stable. 
Therefore, linear perturbation cannot induce the onset of turbulence, and consequently
cannot provide enough viscosity to transport matter inwards therein.

With the application of Magnetorotational Instability 
(MRI; \cite{velikhov,chandra}) to Keplerian disks, Balbus \& Hawley \cite{bh} showed
that initial seed, weak magnetic field can lead to the velocity and magnetic field perturbations
growing exponentially.
Within a few rotation times, such exponential growth could reveal the onset of turbulence. 
Since then, MRI has been a widely accepted mechanism to explain origin of instability and hence transport of 
matter in accretion disks. Note that for flows having strong magnetic fields, where the magnetic field 
is tightly coupled with the flow, MRI is not expected to work.
Hence, it is very clear that 
the MRI is bounded in a small regime of parameter values when field is weak.

It has been well established by several works that transient growth (TG) can reveal nonlinearity and 
transition to turbulence at sub-critical $R_e$ (e.g. \cite{man,amn,chag,yecko,umurhan,avila}). 
Such sub-critical transition to turbulence was invoked to explain colder purely hydrodynamic accretion flows
e.g. quiescent cataclysmic variables, in proto-planetary and star-forming disks, the outer region of disks 
in active galactic nuclei. Baroclinic instability is another plausible source for vigorous turbulence
in colder accretion disks \cite{klar}. Note that while hotter flows are expected to be ionized enough to produce weak 
magnetic fields therein and subsequent MRI, colder flows may remain to be practically neutral in charge
and hence any instability and turbulence therein must be hydrodynamic.
However, in the absence of magnetic effects, the Coriolis force does not allow
any significant TG in accretion disks in three dimensions, independent of $R_e$ 
(see \cite{man}), while in pure two dimensions TG could be large at large $R_e$.
However, a pure two-dimensional flow is a very idealistic case. Nevertheless, in the presence
of magnetic field, even in three dimensions, TG could be very large (Coriolis
effects could not suppress the growth). Hence, in a real three-dimensional flow, it is very important
to explore magnetic TG.

In the present paper, we explore the relative strengths of MRI and TG in magnetized 
accretion flows, in order to explain the generic origin of nonlinearity and
plausible turbulence therein. By TG we precisely mean the short-time scale growth due to 
shearing perturbation waves, producing a peak followed by a dip. By MRI we mean the exponential
growth by static perturbation waves. While TG may reveal nonlinearity in the system, depending 
on $R_e$, amplitude of initial perturbation and its wavevector and background rotational profile of the flow, question 
is, can its growth rate
be fast enough to compete with that of MRI? On the other hand, is there any limitation of MRI,
apart from the fact that MRI does not work at strong magnetic fields? Note that some limitations of 
MRI were already discussed by previous authors \cite{mahajan,umurhan-prl1,umurhan-prl2,avila,pessah},
which then question the origin of viscosity in accretion disks.

We show below that the three-dimensional TG dominates over the growth due to MRI modes 
at large $R_e$, bringing nonlinearity in the flows. 
This is of immense interest, as 
the larger $R_e$ is more plausible in accretion disks.
By comparing modes corresponding to static (original MRI) and shearing (TG) waves,
the growth estimates from static MRI waves have already been argued to be misleading \cite{bhatta}.
Throughout their work, previous authors \cite{bhatta} argued that shearing wave
structures always grow faster over short time scales than
static structures, what we also plan to elaborate here. 
Nevertheless, those authors \cite{bhatta} did not 
explore the length of time over which the short-time growth can persist,
which is very important for revealing non-linearity, which we plan to 
explicitly explore here. We will show below that in a shorter time-scale, TG 
reveals nonlinearity into the system. 

We explicitly calculate the magnetic field strength above which MRI not working.
Moreover, for a fixed perturbation (which might not be either corresponding to the best 
MRI or best TG mode) with finite $R_e$, with the increase of $R_e$, we show that the TG tends 
to bring the nonlinearity in the systems before MRI could do the same, producing a large growth
of perturbation. We notice that above a threshold 
$R_e$, only TG is sufficient to make the system nonlinear at low magnetic field and there 
is no growth at high magnetic fields. Hence, in the regimes of high magnetic field or/and high $R_e$,
MRI is not important at all. 
%In summary, we check the robustness of MRI by varying parameter values 
%and show that MRI is not the sole reason of producing viscosity in accretion disks. 
The working regime of 
MRI is rather much narrower than it is generally thought off.
%and maybe astrophysically implausible. 
%Beyond certain $R_e$, TG overpowers the growth of MRI, leading to the onset of nonlinearity 
%and plausible turbulence and consequently producing viscosity. 
As TG was argued to be plausible source of nonlinearity in cold disks and 
the growth due to MRI is subdominant compared to TG at high $R_e$ in hot disks, 
TG could be argued to be the source of nonlinearity and plausible turbulence and subsequent viscosity, in any
accretion disk.

In the next section, we discuss the perturbation equations describing flows.
Subsequently, we explore total energy growths of perturbations due to TG and MRI
for different parameter values and, furthermore, compare the respective parameter spaces  
for different initial amplitudes of perturbations in \S III and \S IV respectively.
Finally we end with a discussion in \S V.

%\paragraph*{Perturbation equations describing magnetized rotating shear flows.$-$}
\section{Governing equations describing magnetized rotating shear flows in lagrangian coordinates}\label{equation}

Within a local shearing box, in Lagrangian coordinate, the Navier-Stokes, continuity, magnetic induction 
equations and solenoidal condition (for magnetic field) can be written as

%.....................................................................
\begin{eqnarray}
\dot{\bf v}=-\frac{1}{\rho}c_s^2\nabla\rho+\nu\nabla^2{\bf v}+2{\bf v \times \Omega}+\frac{1}{4\pi\rho}{\bf B}\cdot\nabla{\bf B},
\label{ns}
\end{eqnarray}
%.....................................................................
\begin{eqnarray}
\frac{d\rho}{dt}=-\rho\nabla\cdot{\bf v},
\label{cont}
\end{eqnarray}
%.....................................................................
\begin{eqnarray}
\frac{\partial {\bf B}}{\partial t}=\nabla\times({\bf v\times B}),~~
\nabla\cdot{\bf B}=0,
\label{induction}
\end{eqnarray}
%.....................................................................
% \begin{eqnarray}
% \nabla\cdot{\bf B}=0,
% \label{solinoidal}
% \end{eqnarray}
%.....................................................................
when
\begin{eqnarray}
\dot{\bf r}={\bf v}({\bf r}^L),~~
{\bf \nabla} \equiv {\bf \partial r^L \over \partial r}{\bf . \nabla^L},
\label{rl}
\end{eqnarray}
%.....................................................................
% \begin{eqnarray}
% {\bf \nabla} \equiv {\bf
% \partial r^L \over \partial r}{\bf . \nabla^L},
% \label{grad}
% \end{eqnarray}
%......................................................................
where ${\bf v}$ is the velocity vector, ${\bf B}$ the magnetic field, $\nu$ the kinematic coefficient of
viscosity, $c_s$ the sound speed in the shearing box, ${\bf \Omega}$ the angular velocity, ${ \rho}$ 
the density, ${\bf r}$ and ${\bf r}^L$ are the position vectors in Eulerian and Lagrangian coordinates respectively 
\citep{amn}. Note that the contribution of magnetic pressure has been included to the total pressure
in the first term in the right hand side of equation (\ref{ns}). For incompressible flow, equation (\ref{cont}) becomes
%.......................................................................
\begin{eqnarray}
\nabla\cdot{\bf v}=0.
\label{incompress}
\end{eqnarray}
%.......................................................................
Let us define the tensor $\Omega {\mathfrak q}$ which is the minus of the gradient of the unperturbed 
(background) velocity field ${\bf v}_0=(0,-q\Omega x,0)$ as
%.......................................................................
\begin{eqnarray}
\Omega {\mathfrak q} \equiv -\nabla{\bf v}_0=-{\bf (\nabla\Omega)\times R}=\left(
\begin{array}{ccc} 0 & q\Omega & 0 \\ 0 & 0 & 0
\\ 0 & 0 & 0 \end{array} \right);~~ q=-\frac{d\ln \Omega}{d\ln R},
\label{defq}
\end{eqnarray}
%.......................................................................
where ${\bf R}=(R,0,0)$ and $R$ is the distance of the comoving shearing box from the center of the disk and 
$|{\bf \Omega}|=\Omega\propto R^{-q}$ (see \cite{amn} for details). 
Now integrating equation (\ref{rl}), we obtain
%.......................................................................
\begin{eqnarray}
{\bf r}^L = {\bf r} + \Omega t{\bf r} .{\mathfrak q} \Rightarrow {\bf \partial r^L \over \partial r} = {\bf 1} + \Omega t{\mathfrak q},
\label{rlq}
\end{eqnarray}
%.......................................................................
and this gives rise to the relation
%.......................................................................
\begin{eqnarray}
{\bf \nabla} \equiv ({\bf 1} + \Omega t {\mathfrak q}){\bf .\nabla^L}.
\label{gradl}
\end{eqnarray}
%.......................................................................
Since the unperturbed velocity ${\bf v}_0$ has spatial dependence, it has a nonvanishing time derivative in perturbed Lagrangian coordinate. Therefore, we obtain
%.......................................................................
\begin{eqnarray}
\dot{\bf {\delta v}}= \dot{\bf {v}} - \dot{\bf {v}}_0 = \dot{\bf {v}} - {\bf v.\nabla v}_0 = \dot{\bf {v}} 
+ \Omega {\bf v}.{\mathfrak q}.
\label{deltavdot}
\end{eqnarray}
%.......................................................................
Perturbing and linearizing equations (\ref{ns}), (\ref{cont}), (\ref{induction}) and using equation 
(\ref{deltavdot}), we obtain the perturbed Navier-Stokes, continuity, induction equations and solenoidal 
equation for magnetic field in Lagrangian coordinate as
%%......................................................................
%........................................................................
\begin{eqnarray}
\dot{\bf {\delta v}}= -\frac{1}{\rho}c_s^2\nabla\delta\rho+\nu\nabla^2{\bf \delta v}+2{\bf \delta v \times \Omega}+\frac{1}{4\pi\rho}{\bf B}\cdot\nabla{\bf \delta B}+\Omega {\bf \delta v}\cdot{\mathfrak q},
\label{perturbedns}
\end{eqnarray}
%........................................................................
\begin{eqnarray}
\dot{\delta\rho}=-\rho\nabla\cdot{\bf \delta v},
\label{perturbedcont}
\end{eqnarray}
%..........................................................................
\begin{eqnarray}
\dot{\bf {\delta B}}=\nabla\times({\bf v\times\delta B+\delta v\times B})+({\bf v}\cdot\nabla){\bf\delta B},~~~~
\nabla\cdot{\bf \delta B}=0,
\label{perturbedinduction}
\end{eqnarray}
%..........................................................................
where ${\bf \delta v}$, ${\bf \delta B}$ and ${\delta \rho}$ are the velocity, magnetic field vectors and 
the density of perturbation respectively.

We now work with the incompressible approximation, i.e. $\delta \rho \rightarrow 0$ and $c_s^2\rightarrow\infty$, 
assuming $c_s^2\delta \rho$ to be finite and decomposing the 
general linear perturbations into a plane wave form as
%........................................................................
\begin{eqnarray}
{\bf \delta v}, {\bf \delta B}\propto exp(i{\bf k}^L\cdot{\bf r}^L),
\label{perturbation}
\end{eqnarray}
%..........................................................................
when
%........................................................................
\begin{eqnarray}
{\bf k}=(k_x,k_y,k_z)=({\bf 1}+\Omega t{\mathfrak q})\cdot {\bf k}^L=(k_x^L+q\Omega tk_y^L,k_y^L,k_z^L),
\label{wavenumber}
\end{eqnarray}
%..........................................................................
where ${\bf k}$ and ${\bf k}^L$ are the wavevectors in the Eulerian and Lagrangian coordinates respectively. Now using solenoidal condition for magnetic field, incompressibility condition and plane wave solution (\ref{perturbation}), and if we write equations (\ref{perturbedns}) and (\ref{perturbedinduction}) (i.e. Navier-Stokes 
and magnetic induction equations) componentwise, we obtain
%........................................................................
\begin{eqnarray}
\dot{\delta v_x}=-i\frac{1}{\rho}c_s^2\delta\rho(k_x^L+q\Omega tk_y^L)-\nu k^2\delta v_x+2\Omega\delta v_y+\frac{1}{4\pi\rho}i\delta B_x(B_1k_x+B_2k_y+B_3k_z),
\label{nsx}
\end{eqnarray}
%..........................................................................
\begin{eqnarray}
\dot{\delta v_y}=-i\frac{1}{\rho}c_s^2\delta\rho k_y^L-\nu k^2\delta v_y-2\Omega\delta v_x+\Omega q\delta v_x+\frac{1}{4\pi\rho}i\delta B_y(B_1k_x+B_2k_y+B_3k_z),
\label{nsy}
\end{eqnarray}
%..........................................................................
\begin{eqnarray}
\dot{\delta v_z}=-i\frac{1}{\rho}c_s^2\delta\rho k_z^L-\nu k^2\delta v_z+\frac{1}{4\pi\rho}i\delta B_z(B_1k_x+B_2k_y+B_3k_z),
\label{nsz}
\end{eqnarray}
%..........................................................................
\begin{eqnarray}
\dot{\delta B_x}=i\delta v_x(B_1k_x+B_2k_y+B_3k_z),
\label{inductionx}
\end{eqnarray}
%..........................................................................
\begin{eqnarray}
\dot{\delta B_y}=i\delta v_y(B_1k_x+B_2k_y+B_3k_z)-q\Omega\delta B_x,
\label{inductiony}
\end{eqnarray}
%..........................................................................
\begin{eqnarray}
\dot{\delta B_z}=i\delta v_z(B_1k_x+B_2k_y+B_3k_z).
\label{inductionz}
\end{eqnarray}
%..........................................................................

For the convenience of solutions, we further define 
\begin{eqnarray}
\nonumber
\triangle =k_x\delta v_x+k_y\delta v_y,
\zeta =k_x\delta v_y-k_y\delta v_x,
\triangle _B =k_x\delta B_x+k_y\delta B_y,
\zeta _B =k_x\delta B_y-k_y\delta B_x,
\label{newvar}
\end{eqnarray}
and for the plane wave solutions given by equation (\ref{perturbation}), equations (\ref{nsx})-(\ref{inductionz})
 can be recast into
%.......................................................................
\begin{eqnarray}
\left(\begin{array}{c} \dot\triangle \\ \dot\zeta \\ \dot\triangle _B \\ \dot\zeta _B \end{array} \right)
=\left(\begin{array}{cccc} M_{11} & M_{12} & M_{13} & M_{14} \\ M_{21} & M_{22} & M_{23} & M_{24}\\ M_{31} & M_{32} & M_{33} & M_{34}\\M_{41} & M_{42} & M_{43} & M_{44}\end{array} \right) \left(\begin{array}{c} \triangle \\ \zeta \\ \triangle _B \\ \zeta _B \end{array} \right),
\label{newset}
\end{eqnarray}
%...........................................................................
%.......................................................................
where
%.......................................................................
\begin{eqnarray}
\nonumber
&& M_{11}=\frac{-\nu k^4(k_x^2+k_y^2)+2q\Omega k_xk_yk_z^2}{k^2(k_x^2+k_y^2)},~~M_{12}=\frac{2\Omega k_z^2\lbrace k_x^2+(1-q)k_y^2\rbrace}{k^2(k_x^2+k_y^2)},\\
\nonumber
&& M_{13}=\frac{i(B_1k_x+B_2k_y+B_3k_z)}{4\pi\rho},~~M_{14}=0,\\
\nonumber
&& M_{21}=\Omega(q-2),~~M_{22}=-\nu k^2,~~M_{23}=0,~~M_{24}=\frac{i(B_1k_x+B_2k_y+B_3k_z)}{4\pi\rho},\\
\nonumber
&& M_{31}=i(B_1k_x+B_2k_y+B_3k_z),~~M_{32}=0,~~M_{33}=0,~~M_{34}=0,\\
\nonumber
&& M_{41}=0,~~M_{42}=i(B_1k_x+B_2k_y+B_3k_z),~~M_{43}=\frac{q\Omega (k_y^2-k_x^2)}{k_x^2+k_y^2},~~M_{44}=\frac{2q\Omega k_xk_y}{k_x^2+k_y^2}.
\label{coeff}
\end{eqnarray}
%.......................................................................

The assumption of incompressibility is justified as follows. If the wavelength of the velocity 
perturbations is much shorter than the sound horizon for the time of interest (which is 
in the present context the infall time of matter), then the
density perturbations (which is basically the sound waves) reach equilibrium early on, 
which renders effectively
a uniform density during the timescale of interest. For an astrophysical
accretion disk around a black hole, which is either geometrically thin or can be approximated
as a vertically averaged flow, the half-thickness of the disk is comparable
to the sound horizon corresponding to one disk rotation time.
Therefore, as described in previous work (e.g. \cite{amn}), for processes taking longer than
one rotation time, wavelengths shorter than the disk thickness can be approximately treated as
incompressible.

\begin{figure*}
\captionsetup[subfigure]{position=top}
%  \centering
    \begin{tabular}{ll}
 \subfloat[]{\includegraphics[scale=0.7]{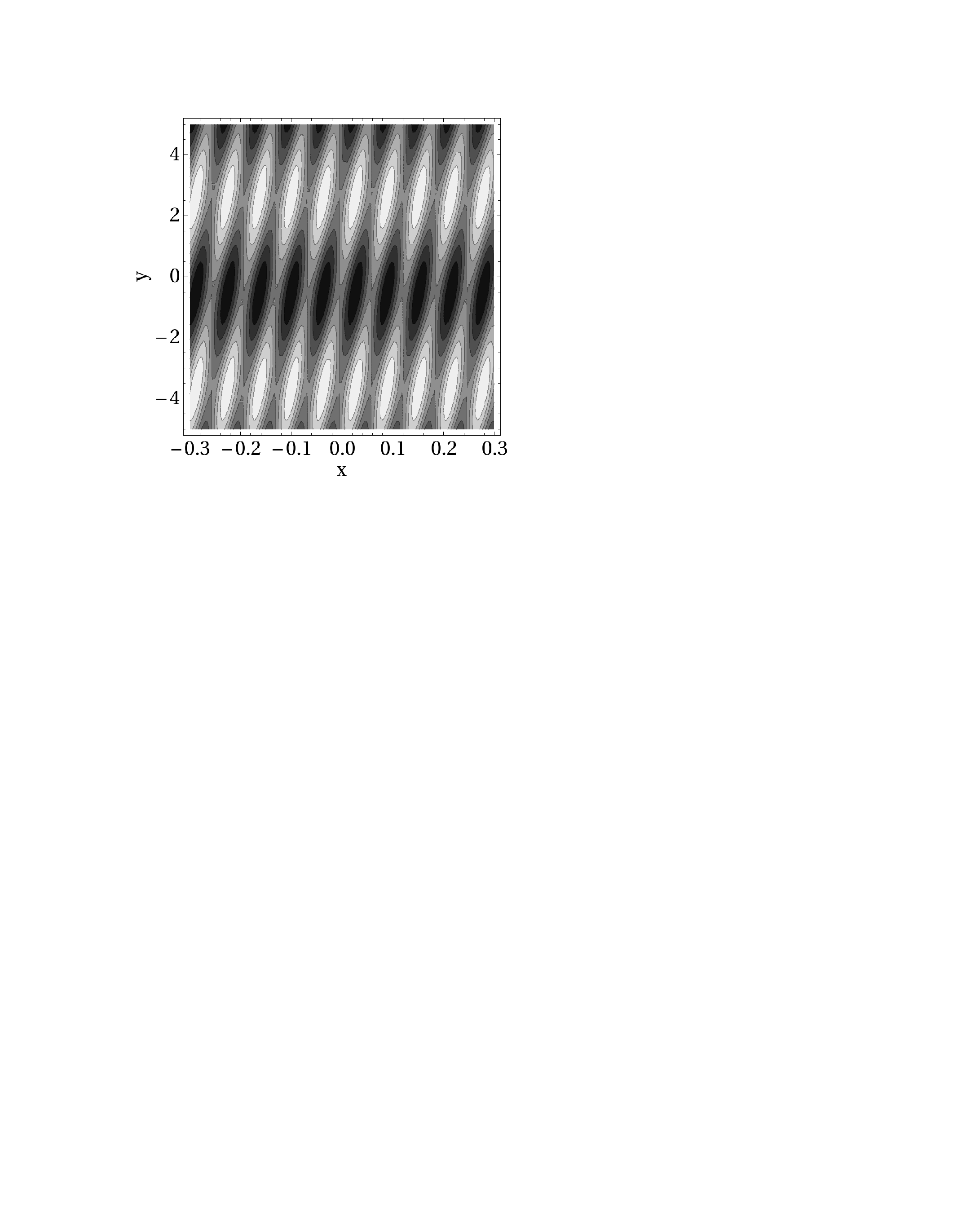}}&
   \subfloat[]{\includegraphics[scale=0.7]{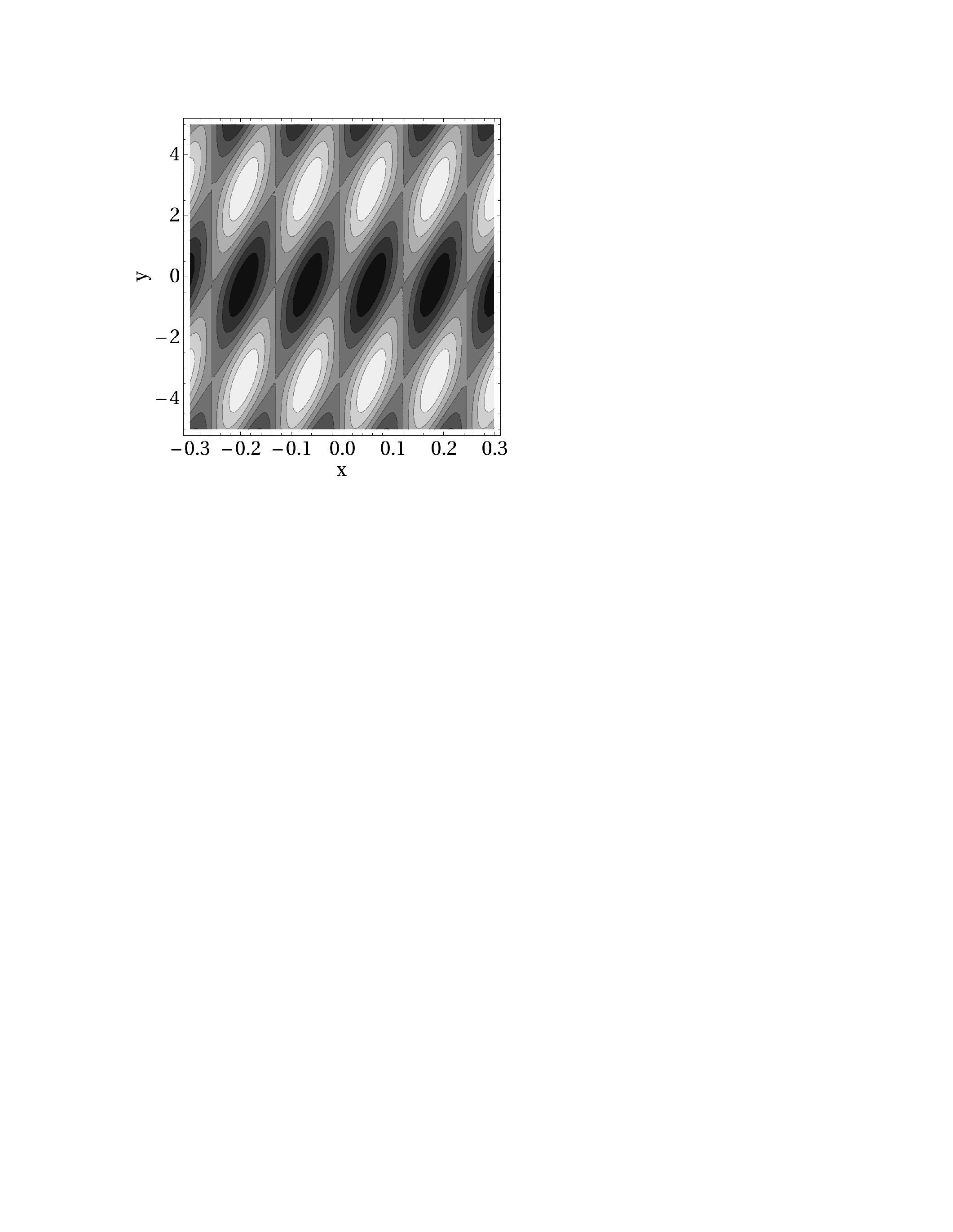}} \\ \\ 
    \subfloat[]{\includegraphics[scale=0.7]{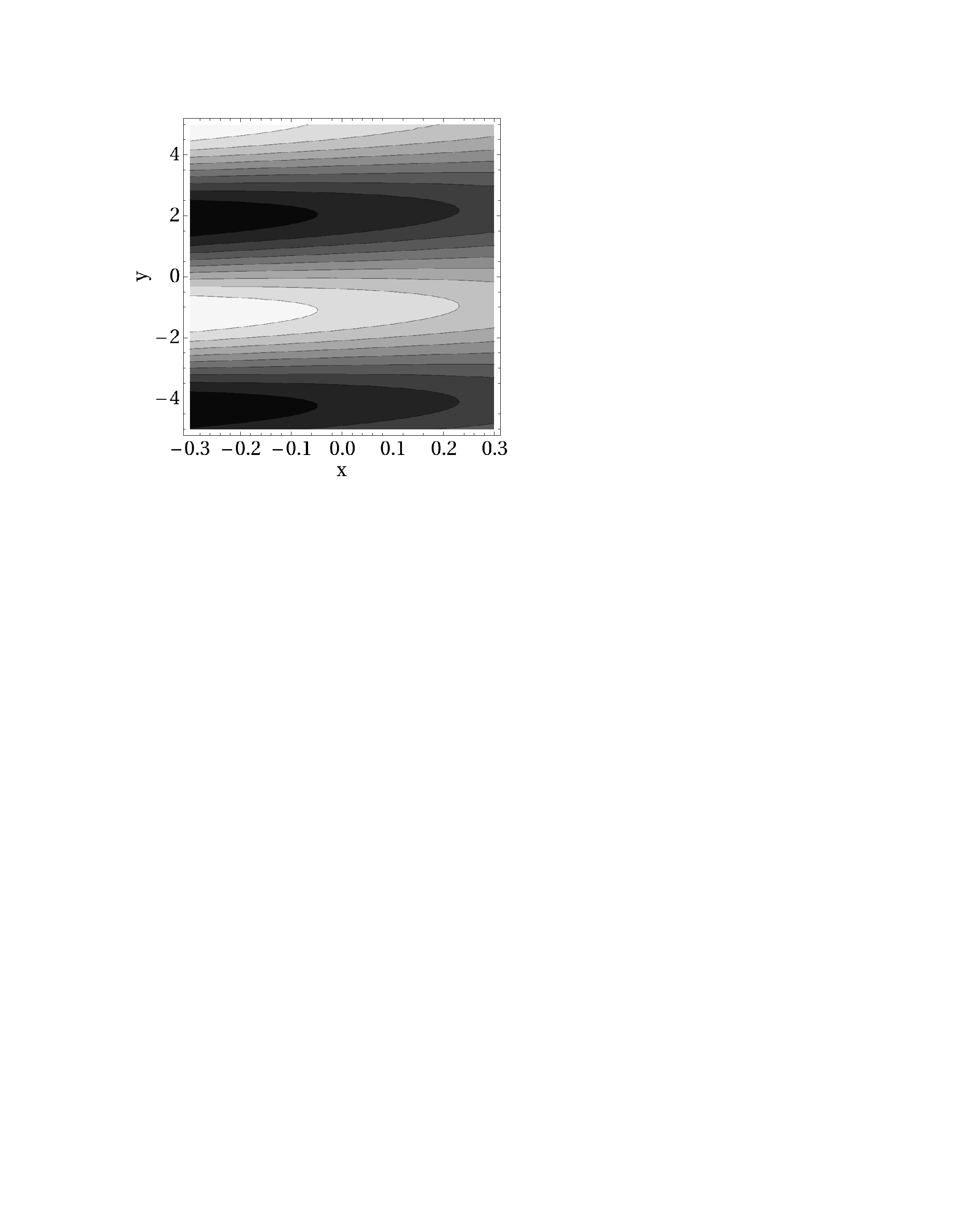}}&
 \subfloat[]{\includegraphics[scale=0.7]{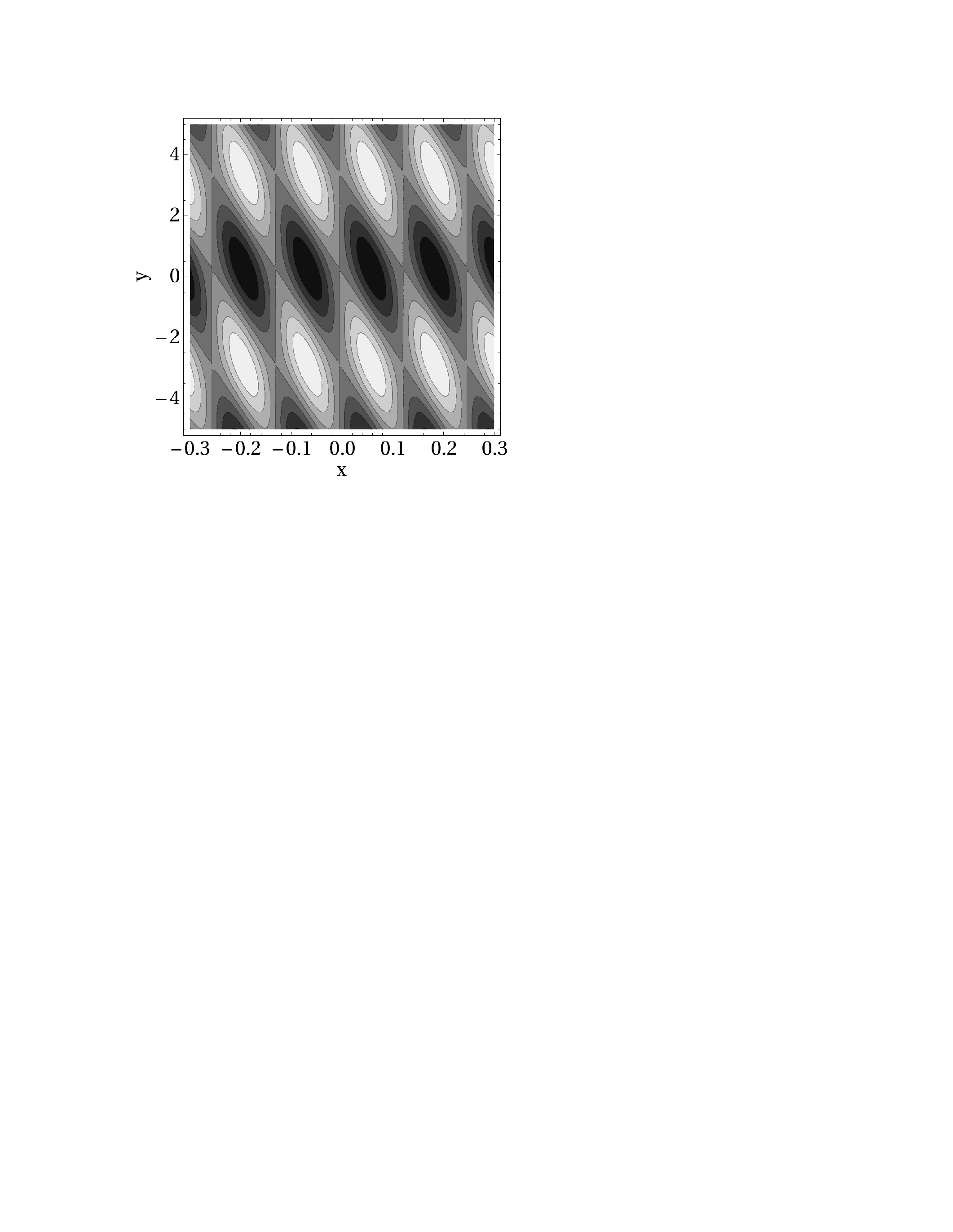}} \\
      \end{tabular}
        \caption{
Development of the perturbed velocity $\delta v_x(x,y)$ as a function of time,
when $R_e=10^6$, $k_y=1$, $z=0$ and $t_{max}$ denotes the time at which growth attains its maximum,
at (a) $t=0$, (b) $t_{max}/2$,
(c) $t_{max}$ and (d) $3t_{max}/2$. The gradual conversion of contour colors from white to black corresponds
to the gradual conversion from positive to negative values of $\delta v_x(x,y)$ respectively.
}
\label{pert}
\end{figure*}

Solving the set of differential equations (\ref{newset}), 
we can calculate {\bf ${\bf\delta v}$, ${\bf\delta B}$} and the energy ${\cal E}$ of the perturbation given by
%.......................................................................
\begin{eqnarray}
{\cal E}\propto \left({\bf\delta v}^2+\frac{{\bf\delta B}^2}{4\pi\rho}\right)=\left(\frac{4\pi\rho\left({\Delta}^2+{\zeta}^2\right)+{\Delta}^2_B+{\zeta} ^2_B}{(k_x+k_yqt\Omega)^2+k^2_y}+\frac{4\pi{\Delta}^2\rho +{\Delta}^2_B}{k^2_z}\right)/8\pi\rho,
\label{totalenergyapp}
\end{eqnarray}
 in terms of new variables.
%.......................................................................
In order to solve the set of equations (\ref{newset}), we have to supply ${\bf\delta B}$ and ${\bf\delta v}$
at $t=0$, i.e. initial perturbation amplitude (IPA). The structure (and evolution) of perturbations are 
similar/same as that found earlier \cite{man,amn,bhatta}.
A sample is shown in Fig. \ref{pert}, demonstrating
how an initial leading wave, with a highly stretched structure, evolves to a spherical wave at the
maximum of TG and furthermore evolves to a trailing wave, during the declining phase
of TG. During this evolution of perturbation,
observing the associated total energy growth of perturbation, we now plan to understand whether the perturbation will sustain 
or not to 
give rise to nonlinearity and plausible turbulence and essentially viscosity to help infall of matter in an accretion disk. 
By a detailed investigation, we can also understand 
%whether the TG itself brings in nonlinearity to the system 
%or it is the MRI mode which does it: what is 
the relative weight between TG and growth due to MRI 
(if at all working) in the time of 
interest. Moreover, we plan to pinpoint the limit of magnetic field strength, above which the
MRI is suppressed (indeed MRI works only for weak magnetic fields).

%\begin{figure*}[h]
%\centering
%\includegraphics[angle=90,width=12.5cm]{lx14052_report_2_b.ps}
%%\includegraphics[angle=0,width=7.5cm]{linearity_comparison_nu_dependent.ps}
%\caption{Development of the perturbed velocity $\delta v_x(x,y)$ as a function of time,
%when $R_e=10^6$, $k_y=1$, $z=0$ at $t=0$ (upper-left), $t_{max}/2$ (upper-right),
%$t_{max}$ (lower-left) and $3t_{max}/2$ (lower-right). Black and white contours correspond 
%to positive and negative values of $\delta v_x(x,y)$ respectively.
%}
%\label{pert}
%\end{figure*}

\begin{figure*}
%\begin{center}
\includegraphics[angle=0,width=8.5cm]{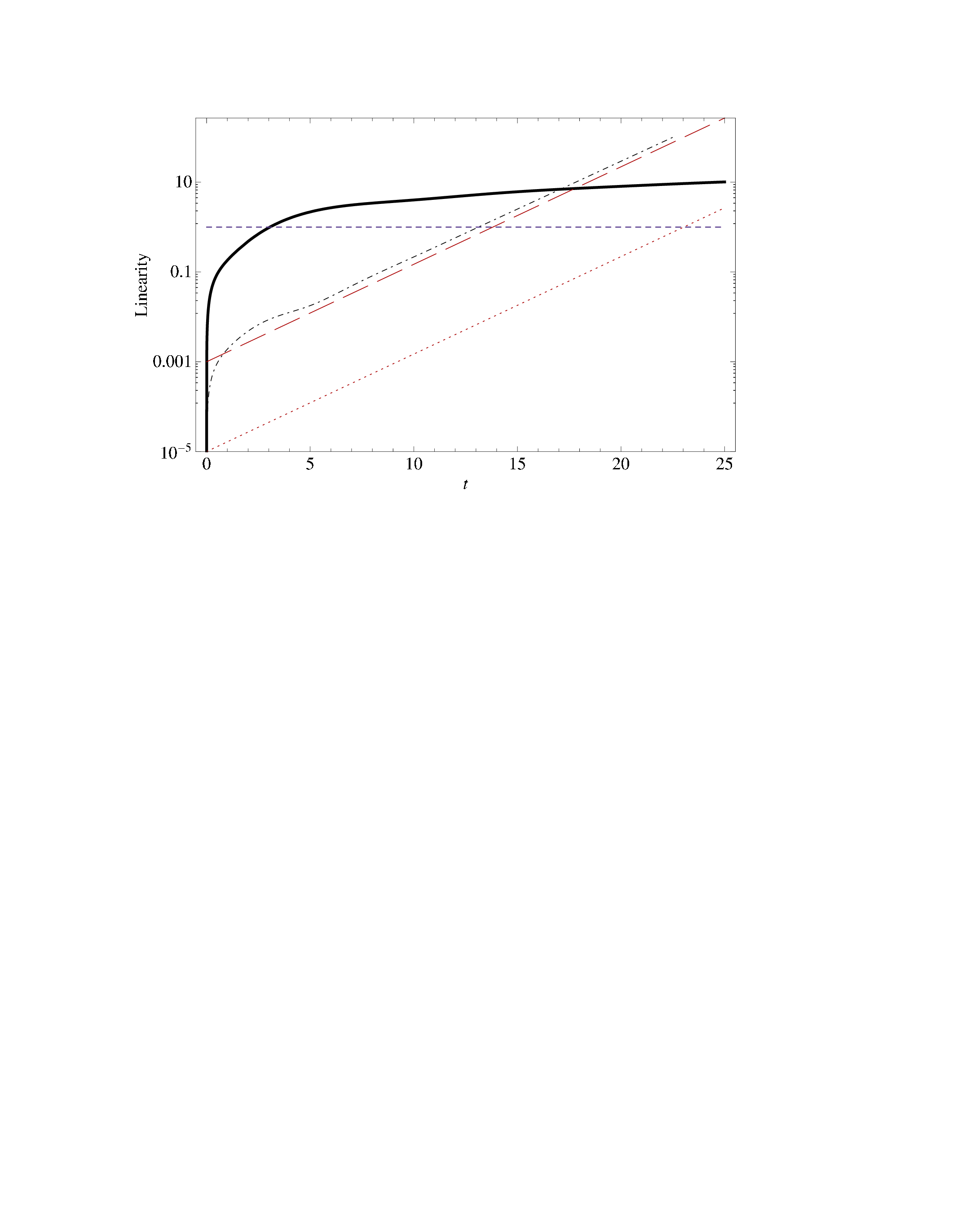}
\caption{(Color online) Nonlinearity via best possible TG and MRI. Thick black line corresponds to
the TG for IPA$=10^{-3}$, $R_e=10^{14}$, $k_{x}^L=-R_e^{1/3}$, $k_y=1$, $k_z=90K_x^L$;
dotdashed black line corresponds to the TG for IPA$=10^{-5}$, $R_e=10^{25}$, $k_{x}^L=-R_e^{1/3}$, 
$k_y=1$, $k_z=90k_x^L$; red longdashed and dotted lines correspond to the best possible MRI starting from 
IPA $=10^{-3}$ and $10^{-5}$ respectively. Dashed horizontal line indicates linearity unity.  
}
\label{lincomnu}
%\end{center}
\end{figure*}

\begin{figure*}
%\begin{center}
\includegraphics[angle=0,width=8.5cm]{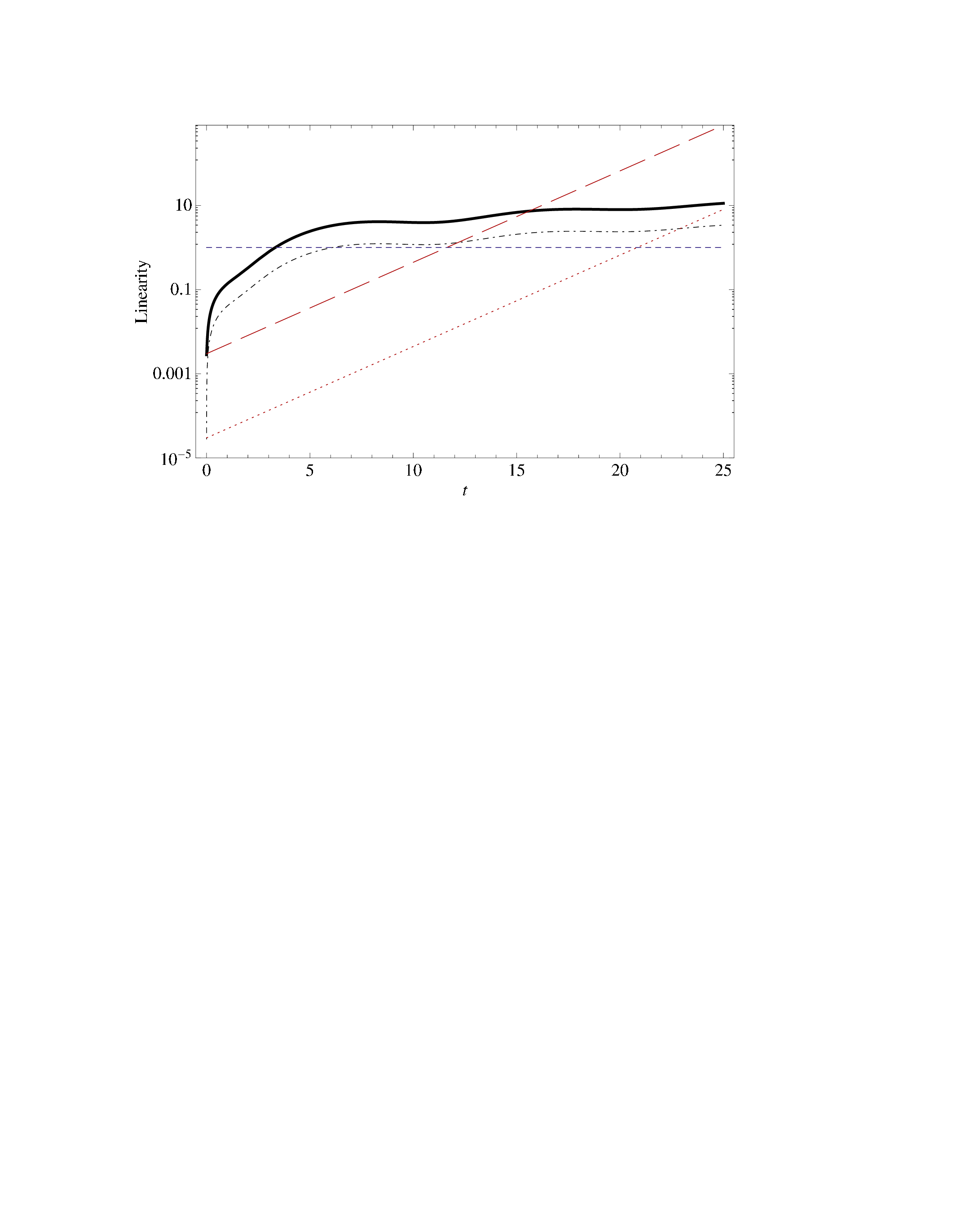}
\caption{Same as Fig. \ref{lincomnu}, but the black thick and dotdashed lines correspond to the TG for 
$k_x^L=1$, $k_y=1$, $k_z=100$, $R_e=10^{12}$. 
and $k_x^L=1$, $k_y=1$, $k_z=3000$, $R_e=10^{12}$ respectively.
}
\label{lincom}
%\end{center}
\end{figure*}

%\paragraph*{Total energy growth of perturbations for different parameter values.$-$}
\section{Total energy growth of perturbations for different parameter values}

The best possible mode for MRI giving rise to the nonlinearity into the system corresponds
to the condition $k_z v_{Az}/\Omega=1$ (when $v_{Az}^2=B_z^2/4\pi\rho$) \cite{bh}. The growth
rate for this fastest exponentially growing mode is $3\Omega/4=3/4q$ (since in dimensionless unit $\Omega=1/q$) \cite{bh,man,bhr}.   
Is there any mode for which TG brings in the nonlinearity 
into the flow (the best possible mode for TG) at a timescale shorter than the rotational time at which the best possible MRI mode brings
in the nonlinearity? 
Note that an approximate emergence of nonlinearity is defined through the measurement
\begin{equation}
{\rm Linearity}=\left(\frac{|{\bf\delta v}|}{|{\bf v}|}+\frac{|{\bf\delta B}|} {|{\bf B}|}\right).
\label{lin}
\end{equation}
When Linearity=1, the system will start becoming nonlinear which will plausibly lead to turbulence.
For a Keplerian disk ($q=3/2$), the best MRI mode brings in the nonlinearity
at the timescales $\sim 14$ and $23$ respectively
for IPAs $=10^{-3}$ and $10^{-5}$ (when $\log(1/{\rm IPA})=3t/4q$). Figure \ref{lincomnu}
shows that indeed there are modes which reveal nonlinearity via TG at around $3$ and 
$13$ rotational times for IPAs $10^{-3}$ and $10^{-5}$ respectively (where $|{\bf \delta v}(0)|/|{\bf v}|, 
|{\bf \delta B}(0)|/|{\bf B}|=\rm IPA$). 
Note from \cite{man} that the maximum TG in two dimensions scales as ${\rm TG_{max}}\sim R_e^{2/3}$ and the 
corresponding time as $R_e^{1/3}$ in pure hydrodynamical disks.
This further corresponds to $k_x^L\sim-R_e^{1/3}$ \cite{bm06}, which reveals
$\rm TG_{max}$  corresponds to the minimum of $k_x$, $k_{x,min}$ \cite{man,amn}.
In the same spirit, $k_x^L$s in Fig. \ref{lincomnu} are chosen to be $-R_e^{1/3}$,
when initial perturbations are highly stretched and nonspherical. Note that although 
such a stretched initial wave vector of perturbation is a special choice which is important for the present 
purpose, nothing prevents them to be arising in nature. Since every perturbation mode is equally 
probable when a system is perturbed (which is indeed the idea behind the choice of the best MRI mode), 
we explore the mode which is growing faster and leading the system to nonlinearity.
In Fig. \ref{lincom}, we relax this $R_e$ dependence of $k_x^L$, but still
obtain the nonlinearity arising at $\sim 3.5$ and $\sim 6$ rotational times for
IPAs $10^{-3}$ and $10^{-5}$ respectively.   
Hence, the full-scale general
hydromagnetic effects giving rise to TG
%, which are based on more general perturbations unlike MRI, 
reveal nonlinearity into the system faster than that the 
MRI does, when MRI itself is uncertain. Once the best TG reveals nonlinearity
before the best MRI would do, the importance of MRI is sluggish in the linear theory. 
Note that our current emphasis, in particular, is the arisal of nonlinearity via either TG or MRI.
However, nonlinearity does not guarantee for the transition to turbulence,
in the physical time scale of accretion. One could argue that MRI modes grow
forever and, hence, the system would have been turbulent at some point, even if the best 
MRI modes are not considered when TG would eventually decay. But the
important fact to notice here is that as soon as the system becomes nonlinear,
MRI (and also TG, if that wins over MRI) is no longer applicable, as the underlying solution 
itself is based on linear theory. On the other hand, the effects due to the best possible MRI mode
should be compared with that of the best possible TG. Such a comparison shows that TG is more powerful 
and is actually responsible for bringing nonlinearity into the systems.

% 
%...............................................................................................
\begin{figure*}
%\begin{center}
\includegraphics[angle=0,width=7.5cm]{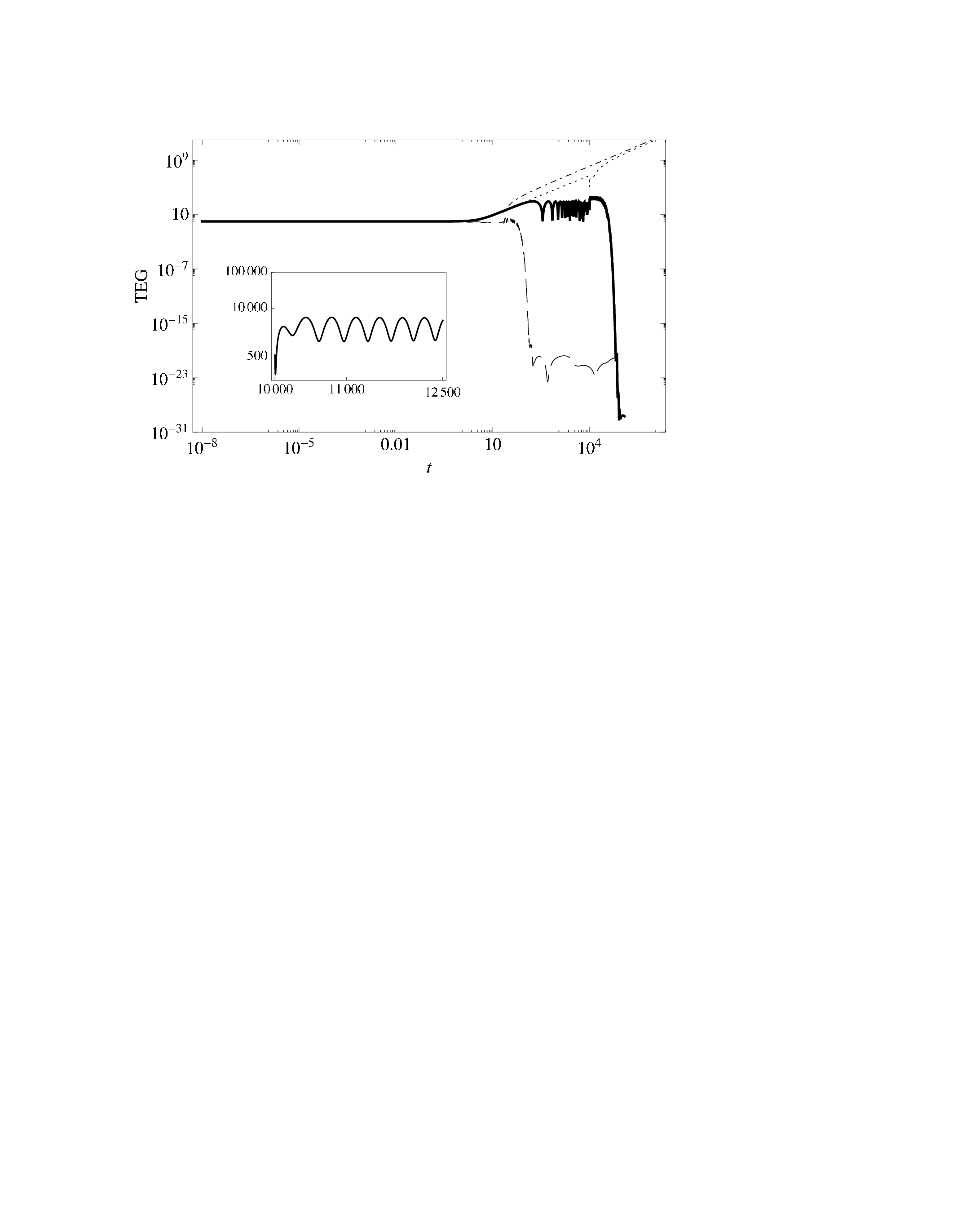}
\caption{Total energy growth for different sets of $R_e$ and ${\bf B}=(0,0,B_3)$: Thick, longdashed, dotted and 
dotdashed lines correspond
to respectively $R_e=10^{12}~{\rm and}~B^2/\rho=10^{-3}$, $R_e=10^{4}~{\rm and}~B^2/\rho=10$, 
$R_e=10^{12}~{\rm and}~B^2/\rho=10^{-20}$ and $R_e=10^{4}~{\rm and}~B^2/\rho=10^{-20}$.
$k_x^L=-R_e^{1/3}$, $k_y=k_z=1$. Inset confirms that the oscillatory zone of 
thick line is continuous and smooth.}
\label{tegall00b3}
%\end{center}
\end{figure*}
%................................................................................................
%...............................................................................................
\begin{figure*}
%\begin{center}
\includegraphics[angle=0,width=8.5cm]{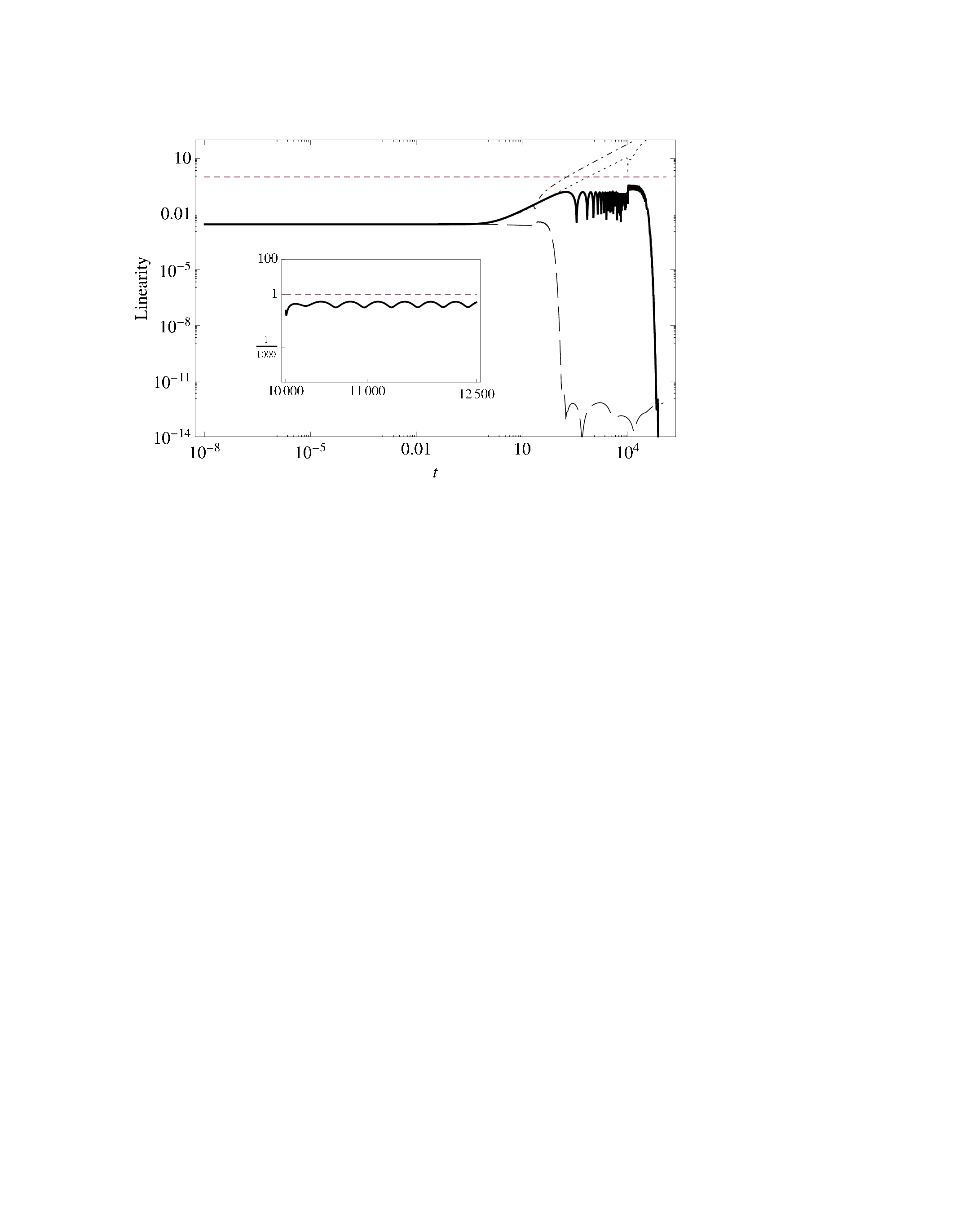}
\caption{Linearity of cases in Fig. \ref{tegall00b3}.
Dashed horizontal line indicates linearity unity.
%The lines are corresponding to $R_e=10^{12} and B^2/\rho=10^{-3}$(Thick), $R_e=10^{4} and B^2/\rho=10$(Dashed), $R_e=10^{12} and B^2/\rho=10^{-20}$(Dotted), $R_e=10^{4} and B^2/\rho=10^{-20}$(Dotdashed).}
Inset confirms that the oscillatory zone of 
thick line is continuous and smooth.}
\label{linearityall00b3}
%\end{center}
\end{figure*}
%................................................................................................
%...............................................................................................
\begin{figure*}
\begin{center}
\includegraphics[angle=0,width=7.9cm]{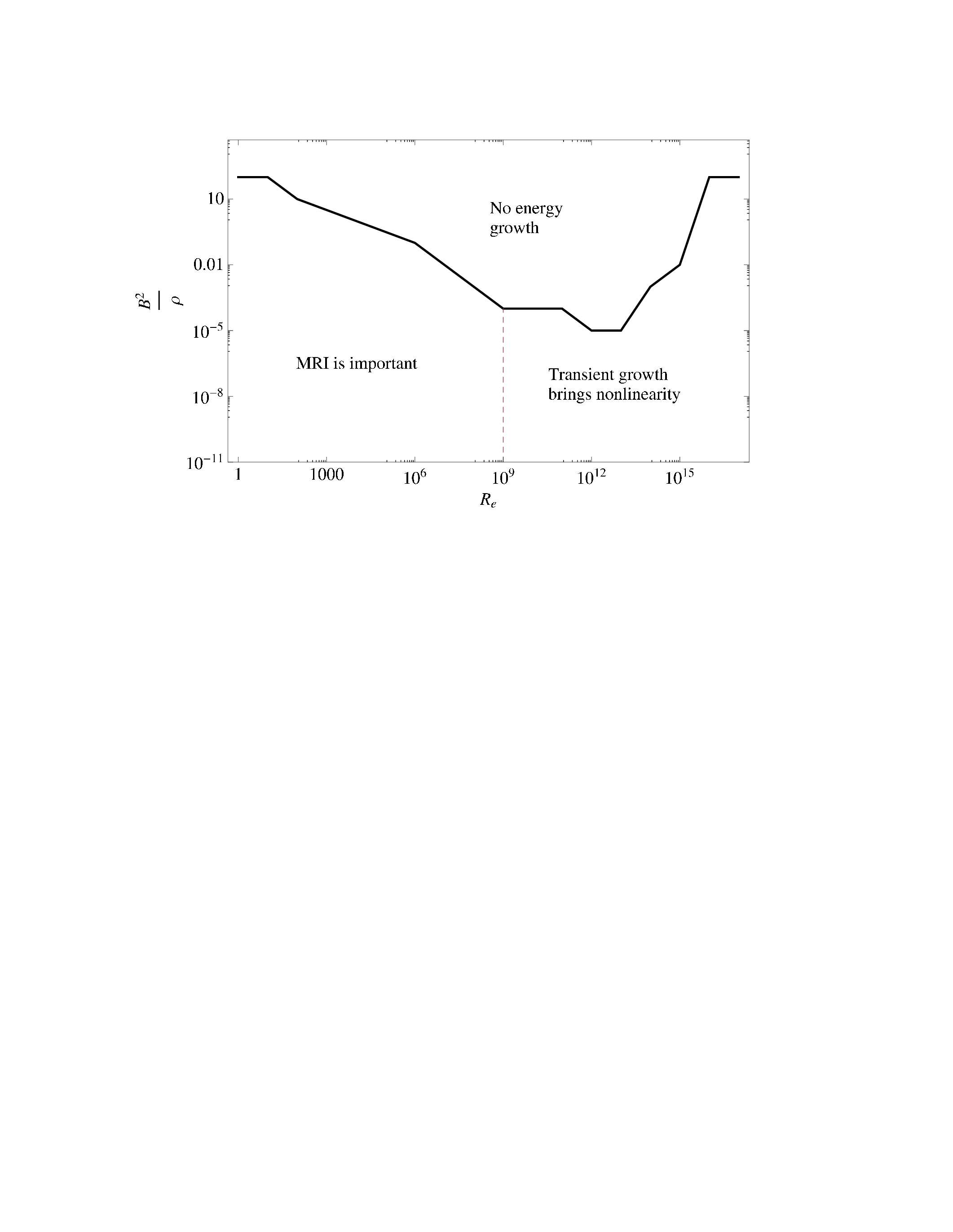}
\caption{Parameter space describing stable and unstable zones, based on the MRI and TG
inactive and active regions, 
when ${\bf B}=(0,0,B_3)$ 
and IPA is $10^{-3}$. The dashed vertical line at $R_e=10^9$ is the threshold 
$R_e$ above which MRI does not work.
$k_x^L=-R_e^{1/3}$, $k_y=k_z=1$.
}
\label{box00b3}
\end{center}
\end{figure*}
%................................................................................................
%...............................................................................................
\begin{figure*}
%\begin{center}
\includegraphics[angle=0,width=8.5cm]{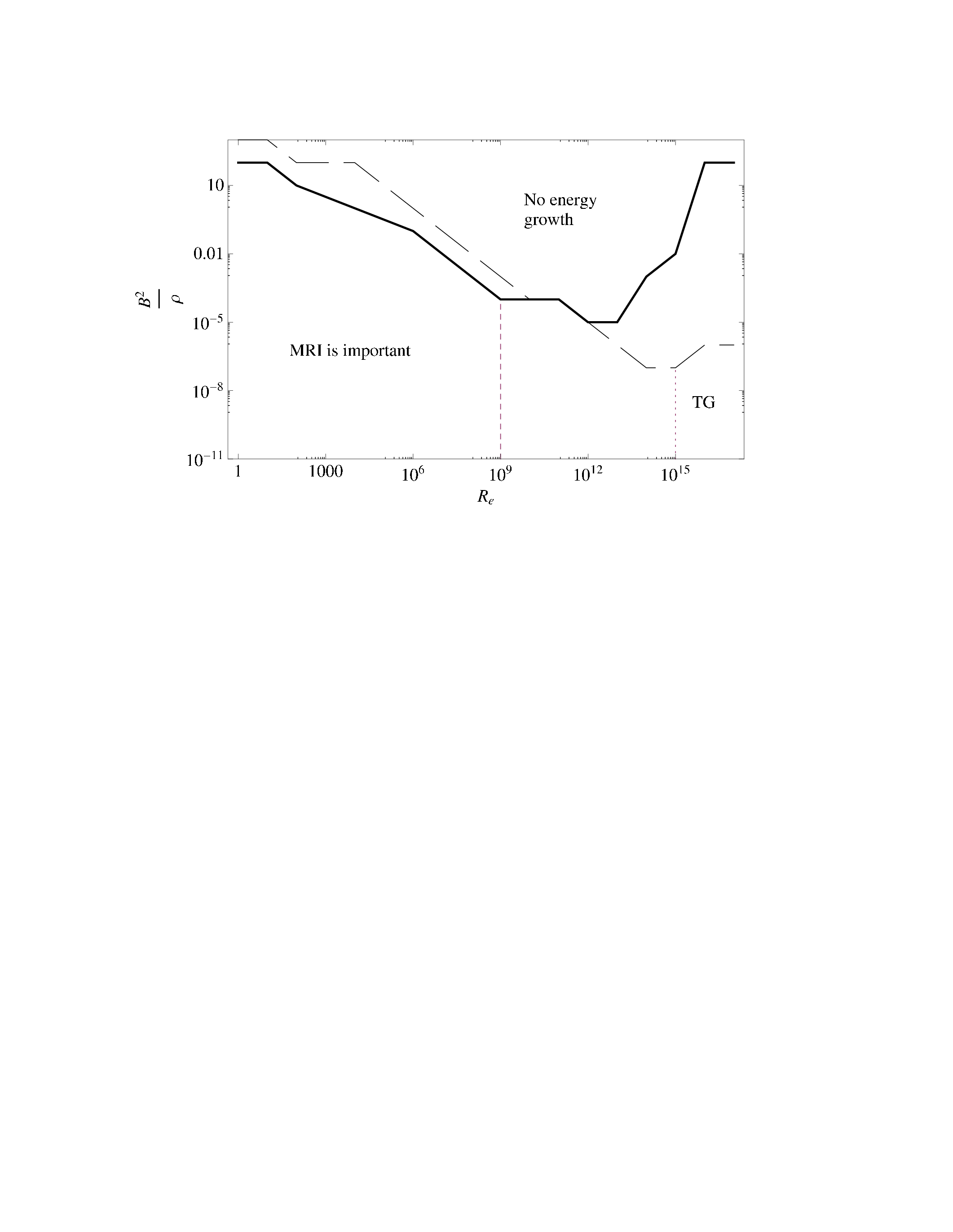}
\caption{Same as Fig. \ref{box00b3} but comparing results with different IPA, when solid and 
longdashed lines are for 
${\rm IPA}=10^{-3}$ and $10^{-5}$ respectively. The dashed and dotted vertical lines at $R_e=10^9$ 
and $10^{15}$ correspond to threshold $R_e$ (as in Fig. \ref{box00b3}) when
${\rm IPA}=10^{-3}$ and $10^{-5}$ respectively.}
\label{box00b3com}
%\end{center}
\end{figure*}
%................................................................................................
Let us move on to the detailed behaviors of TG.
In Fig. \ref{tegall00b3}, we show  energy growths (${\cal E}(t)/{\cal E}(0)$) for four sets of $R_e$
and $B^2/\rho$ (plotted in dimensionless units, based on the dimensions of various quantities of shearing box). 
%from which the strength of magnetic field in Gauss can be calculated as described later in this section). 
We see that for a fixed $R_e$, energy growth of perturbation decays over time if the background magnetic field is 
large (thick and longdashed lines compared to the dotted and dotdashed lines respectively). Figure \ref{linearityall00b3} shows the linearity of respective cases based on equation (\ref{lin}). 
%which is defined as $\left(\frac{|{\bf\delta v}|}{|{\bf v}|}+\frac{|{\bf\delta B}|} {|{\bf B}|}\right)$.
%An approximate emergence of nonlinearity and then plausible
%turbulence in the systems corresponds to the linearity to be unity. 
The most important point to be noted from Fig. \ref{linearityall00b3} is that the case of high $R_e$ 
and low $\bf B$ (dotted line) exhibits nonlinearity
via TG itself, for IPA $\approx 10^{-3}$. Note the clear appearance of TG peak in the 
linearity as well as growth curves at time $t\sim 10^4$. Later, at the trailing phase of TG, growth
further starts increasing due to MRI. However, by this time, the system would already become nonlinear, 
and hence computations of energy growth and that due to MRI based on the linear theory lose their meaning.
However, for a lower $R_e$ for the same $\bf B$, growth due to MRI overpowers TG and nonlinearity
arises via MRI induced growth (dotdashed line).

In order to understand a global picture and the relative powers of TG and MRI, we perform several numerical 
experiments and in Fig. \ref{box00b3} we divide the entire parameter space 
%described by $B^2/\rho$ and $R_e$, 
into three regions: MRI active, TG active and stable (or linear) zones, for a given perturbation 
wavevector. 
Note that for a given $\bf B$, the difference of $\log(R_e)$ between two successive computations is chosen 
to be unity and hence the curve dividing the linearly stable (no energy growth of perturbations) 
and unstable zones does not appear very smooth.
The region left to the solid vertical line exhibits nonlinearity via MRI, while that of right side
corresponds to nonlinearity via TG before MRI could kick in.  

%From the dimensionless value of $B^2/\rho$ we can calculate the strength of magnetic field in Gauss. 
%We can see  in Figure 
%\ref{box00b3} that, for high $R_e$ (higher than $10^9$) and high magnetic fields the total energy growth 
%decays with time. 
Hence, a very important message here is that energy growth
rate due to MRI is faster than TG
only at lower values of $R_e$ and 
it is further suppressed above a certain higher $\bf B$ (when indeed MRI is a weak field effect).
At larger $R_e$, which actually corresponds to astrophysical accretion disks, growth rate due to
TG overpowers that due to MRI.
Although Fig. \ref{box00b3} represents cases corresponding to a vertical background 
magnetic field, we obtain similar trends of results at other background magnetic field geometries and wavevectors,
and hence they are not shown here.

%\paragraph*{Comparison of parameter space for different initial 
%amplitude of perturbations.$-$} 
\section{Comparison of parameter space for different initial amplitude of perturbations}

Figure \ref{box00b3com} compares the parameter spaces, as described in Fig. \ref{box00b3},
for two different IPAs: $10^{-3}$ and $10^{-5}$. 
As IPA decreases, the value of $R_e$ dividing the MRI active and TG active zones, namely $R_{ed}$, increases,
which apparently implies that the MRI active region increases. It also appears that $R_{ed}\sim {\rm IPA}^{-3}$. 
Let us now recall the time scale leading to the system to be nonlinear by the respective growths due to MRI and TG and
estimate if those tally with observation and initial choice of our model.
%in this regime is very high which cannot explain the observed accretion rate. To make this point more clear, we take an example. 
For this purpose, first we fix $R_e$ at $10^{12}$ and take the IPA to be $10^{-3}$, along with sufficiently low $\bf B$
so that the flow is assured to be nonlinear and unstable, in the parameter space described in Figs. \ref{box00b3}
and \ref{box00b3com}. 
In this case, nonlinearity arises in the TG active zone at about $750$ rotation time and corresponding TG is shown in Fig. \ref{linearityall00b3}
(dotted line).  If the shearing box is at $100R_g$ 
away from a $10M_\odot$ black hole, where $R_g$ 
and $M_\odot$ are Schwarzschild radius and Solar mass respectively, then this dimensionless time scale recasts into
$750L/(q\Omega L)=750\sqrt{R^3/GMq^2}\sim 750$ seconds for $q=1.5$ (Keplerian disk), 
when $L$ is the radial width of the shearing box, $G$ the 
Newton's gravitation constant and $M$ the mass of the black hole. Now if we decrease IPA to $10^{-5}$ keeping 
$R_e$ fixed, TG cannot bring nonlinearity any more (however by increasing $R_e$, again TG could
bring nonlinearity), as shown in Fig. \ref{box00b3com}, instead, the nonlinearity 
arises then via MRI modes. However, the time scale for the emergence of nonlinearity, 
in this MRI active process, as shown in Fig. \ref{linearityall00b3}, is approximately $35000$ rotation time
which is $\sim 35000$ seconds (following the same procedure, as used above for the TG active case, 
to convert the dimensionless to dimension-full times). Now we can calculate the radial velocity 
($v_r$) of the Keplerian accretion disk at the location of shearing box for a given
accretion rate $\dot {M}$, say $0.1$ Eddington rate \cite{shakura}, which is supported by observation, given by \cite{shakura}
\begin{eqnarray} 
v_r=2\times 10^6\alpha^{4/5}\left(\frac{\dot{M}}{3\times 10^{-8}M_\odot/{\rm year}}\right)^{2/5}
\left(\frac{M}{M_\odot}\right)^{-1/5}\left(\frac{R}{3R_g}\right)^{-2/5}\left[1-\left(\frac{R}{3R_g}\right)^{-1/2}\right]^{-3/5},
\label{ssvel}
\end{eqnarray} 
when $\alpha$ is the Shakura-Sunyaev viscosity parameter (whose origin is actually aimed here to determine).
It is reasonable to assume that the time required to make the 
flow nonlinear, and hence turbulent which subsequently reveals viscosity, is of the same order as the time 
required by a fluid parcel to cross the length of the shearing box radially ($t_L$) as a result of turbulent viscosity. 
Hence, the product of $v_r$ and $t_L$ should be of the order of width of the shearing box $L$. 
For the above mentioned case of IPA$=10^{-3}$, when nonlinearity is due to TG, we obtain $L=0.1R_g$ from equation 
(\ref{ssvel}), which is highly reasonable for our choice of shearing box 
approximation ($L<<R$). However, for the case of IPA$=10^{-5}$, when nonlinearity is due to MRI,
we obtain $L=10R_g$ which marginally satisfies (or even violates)
the initial choice of a narrow shearing box at $100R_g$. Therefore, 
although smaller IPAs increase MRI active zones, the observed infall cannot be explained by them.
This problem with MRI would appear to be more severe at progressively lower IPAs and TG would
be more important for revealing nonlinearity at progressively higher $R_e$, which are forbidden
for MRI.

%\paragraph*{Discussion and conclusions.$-$} 
\section{Discussion and conclusions}

Let us estimate the maximum $|\bf B|$ in Gauss supporting nonlinearity, as shown by the solid curve(s)
in Figs. \ref{box00b3} and \ref{box00b3com}. We again set the shearing box at $100R_g$ away 
from a $10M_\odot$ black hole. Then we obtain the values of density ($\rho_{100R_g}$) at that location 
to be $\sim 10^{-4}$ gm/cc \cite{shakura}. The background Keplerian 
velocity at that position, for the size of the shearing box, $0.1R_g$, which is consistent with that obtained
for the TG active zone, can be obtained as 
$q\Omega L=q\sqrt{GM/R^3}L\sim 10^6$ cm/sec.
We now consider $R_e=10^{12}$ and, hence, 
from Fig. \ref{box00b3} the corresponding maximum (dimensionless) magnetic field supporting nonlinearity
is given by $B^2/\rho=10^{-5}$. Therefore,
corresponding actual value of magnetic field is $\sqrt{10^{-5}\rho_{100R_g}(q\Omega L)^2}\sim 30$ Gauss. 
This means, the flow with $R_e=10^{12}$ and 
$|{\bf B}|>30$ 
Gauss, the energy growth of perturbation will decay over time, but for $|{\bf B}|\leqslant 30$ Gauss, TG 
will be sufficient enough to bring nonlinearity in the system, however, still not requiring any growth due to MRI. 
>From Fig. \ref{box00b3}, 
it is clear that MRI is only important whenever $R_e < 10^9$, whereas for $R_e \geqslant 10^9$, which is the 
favorable zone of $R_e$ for accretion disks, magnetic TG is important than MRI.

In short, we have 
calculated the magnetic field strengths for different $R_e$s above which the system will be stable under 
linear perturbation and an upper bound of $R_e$ above which either the system is stable under linear perturbation 
(for high magnetic field strength) or reaches nonlinear regime (for low magnetic field) through magnetic TG. 
In one line, MRI is not the sole mechanism to make accretion disk unstable, there is a big kingdom where TG rules,
and explanation of accretion solely via MRI is misleading.\\

The authors would like to thank Prateek Sharma for illuminating discussion.
B.M. acknowledges partial support through the research grant provided by 
Indian Space Research Organization of Ref. No. ISRO/RES/2/367/10-11.


\begin{thebibliography}{73}

\bibitem{pringle}
 J.E. Pringle, ARA\&A \textbf{19,} 137 (1981).

\bibitem{bmplb}
B. Mukhopadhyay, Phys. Lett. B \textbf{721,} 151 (2013).

\bibitem{velikhov}
E. Velikhov, J. Exp. Theor. Phys. \textbf{36,} 1398 (1959).

\bibitem{chandra}
S. Chandrasekhar, Proc. Nat. Acad. Sci. (USA) \textbf{46,} 253 (1960).

\bibitem{bh}
 S.A. Balbus, and J.F. Hawley, Astrophys. J. \textbf{376,} 214 (1991).

\bibitem{man}
B. Mukhopadhyay, N. Afshordi, and R. Narayan, Astrophys. J. \textbf{629,} 383 (2005).

\bibitem{amn}
N. Afshordi, B. Mukhopadhyay, and R. Narayan, Astrophys. J. \textbf{629,} 373 (2005).

\bibitem{chag}
G.D. Chagelishvili, J.-P. Zahn, A.G. Tevzadze, J.G. Lominadze, Astron. Astrophys. \textbf{402,} 401 (2003).

\bibitem{yecko}
P.A. Yecko, Astron. Astrophys. \textbf{425,} 385 (2004).

\bibitem{umurhan}
 O.M. Umurhan, and O. Regev, Astron. Astrophys. \textbf{427,} 855 (2004).

\bibitem{avila}
M. Avila, Phys. Rev. Lett. \textbf{108,} 124501 (2012).

\bibitem{klar} H.H. Klahr, and P. Bodenheimer, Astrophys. J. \textbf{582,} 869 (2003).

\bibitem{mahajan}
S.M. Mahajan, V. Krishan, Astrophys. J. \textbf{682,} 602-607 (2008).

\bibitem{umurhan-prl1}
O.M. Umurhan, K. Menou, and O. Regev, Phys. Rev. Lett. \textbf{98,} 034501 (2007).

\bibitem{umurhan-prl2}
E. Liverts, Y. Shtemler, M. Mond, O.M. Umurhan, and D.V. Bisikalo, Phys. Rev. Lett.
\textbf{109,} 224501 (2012).

\bibitem{pessah}
M.E. Pessah, and C. Chan, Astrophys. J. \textbf{751,} 48 (2012).

\bibitem{bhatta}
J. Squire and A. Bhattacharjee, Phys. Rev. Lett. \textbf{113,} 025006 (2014).

 \bibitem{bhr}
S.A. Balbus, and J.F. Hawley, Rev. Mod. Phys. \textbf{70,} 1 (1998).

 \bibitem{bm06}
B. Mukhopadhyay, Astrophys. J. \textbf{653,} 503 (2006).

\bibitem{shakura}
 N.I. Shakura, and R.A. Sunyaev, Astron. Astrophys. \textbf{24,} 337 (1973).


\end{thebibliography}
\end{document}